# Compact Dual-Polarized Vivaldi Antenna for Ground Penetrating Radar (GPR) Application

Hai-Han Sun, Yee Hui Lee, Abdülkadir C. Yücel
School of Electrical & Electronic Engineering
Nanyang Technological University, Singapore
haihan.sun@ntu.edu.sg, eyhlee@ntu.edu.sg,
acyucel@ntu.edu.sg

Genevieve Ow, Mohamed Lokman Mohd Yusof
Centre for Urban Greenery & Ecology
National Parks Board, Singapore
GENEVIEVE_OW@nparks.gov.sg,
Mohamed_Lokman_Mohd_Yusof@nparks.gov.sg

*Abstract*—In this paper, a compact dual-polarized Vivaldi antenna is presented. Four Vivaldi elements are used as radiators, and are positioned obliquely and connected in a horn shape. By exciting two sets of elements, two orthogonally polarized radiations can be achieved. The dual-polarized antenna features a low operating frequency band with a wide bandwidth, high port isolation, good directional radiation performance, and a very compact size, making it highly suitable for the ground penetrating radar (GPR) application.

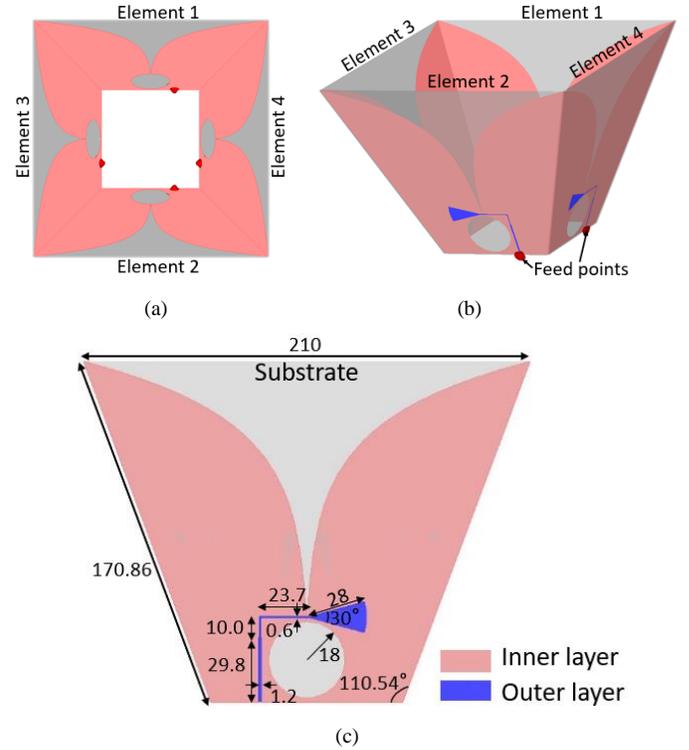

Fig. 1 Configuration of the dual-polarized Vivaldi antenna. (a) Top view, (b) perspective view, and (c) front view of each Vivaldi element with optimized parameters. (Dimensions in mm unless specified.)

## I. INTRODUCTION

Ground Penetrating Radar (GPR) has been widely used as a non-destructive method to detect or image subsurface objects and structures. Antennas are an essential component in the GPR system. It is often a challenge to design an antenna to meet all the specifications. These specifications include a low operating frequency range with an ultrawide bandwidth to guarantee the penetration depth and resolution. Small electrical size with high gain and directional radiation performance is especially important for a light-weight and portable GPR system. As the investigated objects may depolarize electromagnetic waves incident upon them, dual-polarized radiation is desired to collect more information and improve the detection capability [1].

Vivaldi antenna has been a popular choice for GPR applications with its wide bandwidth, good directional performances, and ease of fabrication. Dual-polarized Vivaldi antennas are constructed in two ways. The Vivaldi elements are either orthogonally arranged along their outer edges [2], [3], or orthogonally inserted into each other, forming a cross shape [4]-[6]. Wide operating bandwidth with moderate port isolation level and cross-polarization discrimination (XPD) have been reported in the literature [2]-[6]. However, most of these antennas have a large electrical size, which may not be handy when mounted in a confined space. More research needs to be carried out to realize a compact dual-polarized Vivaldi antenna with improved dual-polarized performance characteristics.

In this work, we present a novel shared-aperture dual-polarized Vivaldi antenna. The antenna has a broad operating band from 0.4 GHz to 2.5 GHz with port isolation of more than 60 dB, good directive radiation with high-gain performance. With the shared-aperture configuration, the antenna has a very compact size of $0.28\,\lambda \times 0.28\,\lambda \times 0.23\,\lambda$ ($\lambda$ is the wavelength at 0.4 GHz), making it highly suitable for a portable GPR system.

## II. CONFIGURATION OF THE DUAL-POLARIZED VIVALDI ANTENNA

The configuration of the dual-polarized Vivaldi antenna is shown in Fig. 1. Four obliquely positioned Vivaldi antenna elements are connected in a horn shape. Two parallel elements are excited simultaneously for one polarization, so the two orthogonal polarizations are realized by exciting elements 1 and 2, and elements 3 and 4, respectively. Each Vivaldi element is fed by an optimized microstrip-slot balun. A detailed view of the Vivaldi element with optimized dimensions is shown in Fig. 1(c). The substrate used in this work has a dielectric constant of 4.4, a loss tangent of 0.0025, and a thickness of 1.0 mm.

This work is funded by National Parks Board, Singapore.

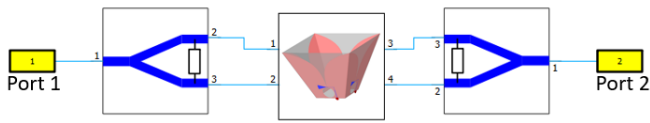

Fig. 2 Illustration of the antenna excitation.

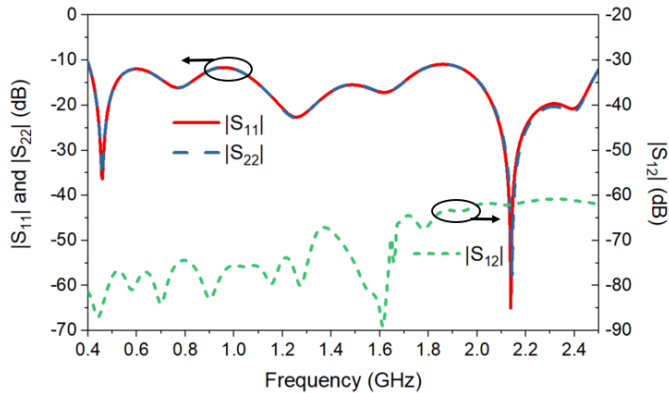

Fig. 3 Simulated S-parameters of the dual-polarized Vivaldi antenna.

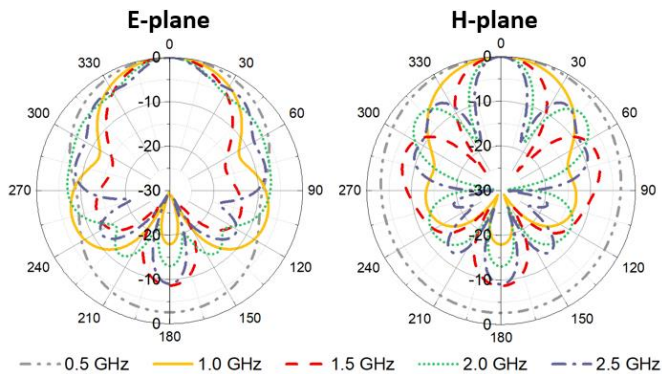

Fig. 4 Normalized radiation patterns in the E- and H-planes of port 1.

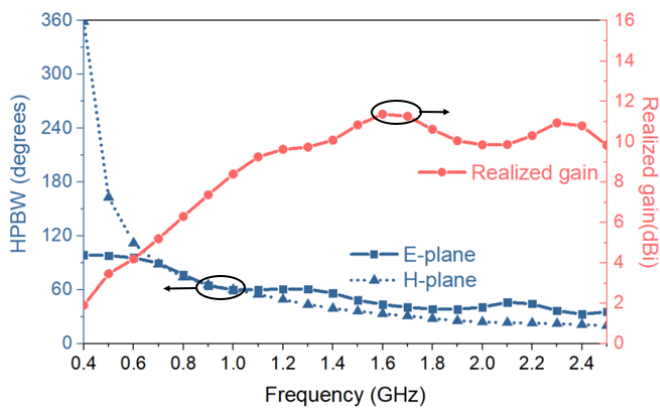

Fig. 5 Simulated HPBW in the E- and H-planes and realized gain result.

## III. ANTENNA'S PERFORMANCE

The antenna is simulated using CST Studio Suite 2019. The two sets of elements are fed by two ideal 3 dB power dividers, as shown in Fig. 2. Ports 1 and 2 excite the two orthogonally polarized radiations respectively. Fig. 3 shows the S-parameters of the antenna. The antenna has a wide operating bandwidth of 145% from 0.4 GHz to 2.53 GHz with reflection coefficients less than -10 dB. The $|S_{12}|$ is less than -60 dB over this band, which represents a very high isolation level between the orthogonal ports. Ports 1 and 2 yield identical radiation performance due to the symmetrical arrangement of the antenna structure. The E-plane and H-plane radiation patterns of port 1 at 0.5 GHz, 1.0 GHz, 1.5 GHz, 2.0 GHz, and 2.5 GHz are presented in Fig. 4. The half-power beamwidths (HPBW) of E-plane and H-plane patterns and the realized gain are shown in Fig. 5. The beams in both E- and H-planes become narrower and the realized gain increases with frequency. The maximum gain reaches 11.4 dBi at 1.6 GHz. Fig. 4 and Fig. 5 also show that directional radiation patterns with similar HPBWs in both E- and H- planes are realized from 0.6 GHz onwards. The simulated XPD is more than 20 dB in the main beam coverage across the band. The total efficiency of the antenna is more than 90%. The antenna has a very compact size of 210 mm × 210 mm × 170 mm, which is equivalent to 0.28 λ × 0.28 λ × 0.23 λ at 0.4 GHz.

## IV. CONCLUSION

In this paper, we present a dual-polarized Vivaldi antenna with a very compact size for the GPR application. Four connected Vivaldi elements are divided into two sets for dual-orthogonal-polarized radiation. The antenna has a -10 dB impedance bandwidth of 145% with port isolation higher than 60 dB. Directional radiation with high XPD and high gain performance characteristics are achieved. Moreover, the antenna features a very compact size of 0.28 λ × 0.28 λ × 0.23 λ at 0.4 GHz. The antenna is particularly suitable for portable GPR systems, and can be further implemented in antenna arrays. The experimental results and the detection capability of this antenna will be presented at the conference.

## REFERENCES

[1] A. Simi, G. Manacorda, and A. Benedetto, "Bridge deck survey with high resolution ground penetrating radar," in *Proc. 14th Int. Conf. Ground Penetrating Radar (GPR)*, Jun, 2012, pp. 489–495.

[2] M. Wajih Elsallal and J. C. Mather, "An ultra-thin, decade (10:1) bandwidth, modular "BAVA" array with low cross-polarization," in *Proc. IEEE Int. Symp. Antennas Propag. (APSURSI)*, Jul. 3–8, 2011, pp. 1980–1983.

[3] R. W. Kindt and W. R. Pickles, "Ultrawideband all-metal flared-notch array radiator," *IEEE Trans. Antennas Propag.*, vol. 58, no. 11, pp. 3568–3575, Nov. 2010.

[4] G. Adamiuk, T. Zwick, and W. Wiesbeck, "Compact, dual-polarized UWB-antenna, embedded in a dielectric," *IEEE Trans. Antennas Propag.*, vol. 58, no. 2, pp. 279–286, Feb. 2010.

[5] X. Han, L. Juan, C. Changjuan, and Y. Lin, "UWB dual-polarized Vivaldi antenna with high gain," in *Proc. Int. Conf. Microw. Millim. Wave Technol. (ICMMT)*, May, 2012, pp. 1–4.

[6] M. Sonkki, D. Sánchez-Escuderos, V. Hovinen, E. T. Salonen, and M. Ferrando-Bataller, "Wideband dual-polarized cross-shaped Vivaldi antenna," *IEEE Trans. Antennas Propag.*, vol. 63, no. 6, pp. 2813–2819, Jun. 2015.